\documentclass[10pt,twocolumn,letterpaper]{article}
\pdfoutput=1
\usepackage{cvpr}
\usepackage{times}
\usepackage{epsfig}
\usepackage{graphicx}
\usepackage{amsmath}
\usepackage{amssymb}
\usepackage{array}
\usepackage[linesnumbered,boxed,vlined]{algorithm2e}
\SetKwInput{KwInput}{Input}
\SetKwInput{KwOutput}{Output}
\usepackage[export]{adjustbox}
\usepackage{enumerate}
\usepackage{enumitem}
\usepackage{url}
\usepackage{caption}
\usepackage{subcaption}
\usepackage{xcolor}
\usepackage[export]{adjustbox}
\usepackage{breqn}
\usepackage{adjustbox}
\usepackage{pifont}
\newcommand{\cmark}{\ding{52}}%
\newcommand{\xmark}{\ding{54}}%

\usepackage[breaklinks=true,colorlinks,bookmarks=false]{hyperref}

\cvprfinalcopy 

\def\cvprPaperID{12} 

\ifcvprfinal\fi
\begin{document}

\title{Multi-level Encoder-Decoder Architectures for Image Restoration}

\author{Indra Deep Mastan and Shanmuganathan Raman\\
Indian Institute of Technology Gandhinagar\\
Gandhinagar, Gujarat, India\\
{\tt\small \{indra.mastan, shanmuga\}@iitgn.ac.in}}

\maketitle
\thispagestyle{empty}

\begin{abstract}
Many real-world solutions for image restoration are learning-free and based on handcrafted image priors such as self-similarity. Recently, deep-learning methods that use training data, have achieved state-of-the-art results in various image restoration tasks (e.g., super-resolution and inpainting). Ulyanov \textit{et al.} bridge the gap between these two families of methods in  \cite{Ulyanov2018CVPR}. They have shown that learning-free methods perform close to the state-of-the-art learning-based methods ($\approx$ 1 PSNR). Their approach benefits from the encoder-decoder network ($ed$). 

In this paper, we propose a framework based on the multi-level extensions of the encoder-decoder network ($med$) to investigate interesting aspects of the relationship between image restoration and network construction independent of learning. Our framework allows various network structures by modifying the following network components: skip links, cascading of the network input into intermediate layers, a composition of the encoder-decoder subnetworks, and network depth. These handcrafted network structures illustrate how the construction of untrained networks influence the following image restoration tasks: denoising, super-resolution, and inpainting. We also demonstrate image reconstruction using flash and no-flash image pairs. We provide performance comparisons with the state-of-the-art methods for all the restoration tasks above. The additional results are available on the \href{https://indradeepmastan.github.io/image-restoration/}{project page}.
\end{abstract}

\section{Introduction}
Image restoration is an \textit{ill-posed} problem which aims to recover an image given its corrupted observation (e.g., denoising  \cite{zhu2016noise,burger2012image,vincent2008extracting}, super-resolution \cite{ledig2016photo,sajjadi2016enhancenet,bhowmik2018training}, and inpainting \cite{yeh2017semantic,yang2017high,xie2012image}). 
Corruption may occur due to noise, camera shake, and due to the fact that the picture was taken in rain or underwater \cite{li2018learning}. Image restoration methods could be mainly classified into two types - \emph{traditional} methods and  \emph{deep-learning} (DL) methods.  Traditional methods include  spatial filtering methods (e.g., bilateral filters \cite{tomasi1998bilateral}, non-local means \cite{buades2005non}), wavelet transform based methods \cite{chang2000adaptive}, and dictionary learning and sparse coding \cite{mairal2010online,zeyde2010single}. DL methods generally include a neural network to learn image prior from the training samples (learning-based\footnote{The learning refers to training the network on the collection of images and learning-free refers to the methods which do not use training data.}) for restoration, where the training samples contain paired examples of corrupted and high-quality images. 

Traditional methods are generally faster and comparatively less cumbersome to implement, \textit{e.g.,} filtering approaches \cite{duchon1979lanczos}. Whereas DL methods could be tricky to implement. For example, methods based on adversarial loss require training of two separate networks, namely a generator and a discriminator \cite{ledig2016photo}. Moreover, DL methods output photo-realistic images with finer details of features due to the image prior being captured by feature learning on a collection of images \cite{ledig2016photo,zhang2017learning,bigdeli2017image}. 

Representation learning from images gives insight into the image statistics captured by the network. The main idea is to perform various image restoration tasks to learn a better image prior \cite{kim2018learning}. However, it is focused on the learning-based setting  \cite{bau2017network}. There are fewer studies that directly investigate the image prior captured by the neural network without using any training datasets. Ulyanov \textit{et al.} first conducted the studies to achieve image restoration without using a training sample (learning-free) \cite{Ulyanov2018CVPR}. This paper focuses on the research thread mentioned above. Our work  combine the ideas of traditional methods and the DL approaches similar to \cite{gandelsman2018double, shocher2018internal, lefkimmiatis2017non,yang2018bm3d,Ulyanov2018CVPR}.

Our ablation study shows how the structure of the untrained network influences the quality of image restoration achieved by them. For example, inpainting of a large missing region is qualitatively better-achieved using an encoder-decoder network \emph{without} skip connections, whereas super-resolution is better-achieved \emph{with} skip links (Fig.~\ref{fig: skip1} and Fig.~\ref{fig: skip2}). 

We have performed extensive experiments on various handcrafted network architectures obtained by modifying the network components.  We focus on the following network components: \textit{depth} of the network, \textit{skip connections}, cascading of the network input into intermediate layers (\textit{cascade}), and composition of the encoder-decoder subnetworks (\textit{composition}). We show how each of the above network components affects image restoration. For example, we show how the performance of denoising gets affected when we increase the depth of the network (Fig.~\ref{fig: ed-depthEffects}). 

We have formulated a framework called multi-level encoder decoder ($med$) that models various handcrafted network architectures. An instance from our framework $med$ is a composition of three encoder-decoder networks (Fig.~\ref{fig: MEDS}). The multi-level extension of encoder-decoder is motivated to exploit the re-occurrence of an image patch at different resolutions. We show our analysis using six different network instances of $med$ (Table~\ref{tab: medClass}). These handcrafted network architectures help us develop insight into how the network construction influences image restoration (Fig.~\ref{fig: ed-depthEffects}, Fig.~\ref{fig: objectRemove-cascade}, Fig.~\ref{fig: textRemove-composition},  Fig.~\ref{fig: skip1}, and Fig.~\ref{fig: skip2}). The key idea is to iteratively minimize the loss between the network output and the corrupted image to implicitly capture the image prior in the network. 

There is an inherent contrast in our objectives. On the one hand, we aim to experiment with various high capacity networks to show the relation between image restoration and network construction. The higher depth allows more network components and various network structures for the analysis of the image prior. On the other hand, the high capacity network should not negatively influence the quality of image restoration. This is due to the fact that the higher depth network suffers from the \emph{vanishing gradients} problem \cite{mao2016image, srivastava2015training}. One option is to use skip links to propagate the gradients and feed the image features from the intermediate layers to the last layers of the network \cite{mao2016image}.  Our main contributions are summarized as follows.
\begin{itemize}[leftmargin=*]
\item To the best of our knowledge, this is the first study of a multi-level encoder-decoder framework ($med$) designed to illustrate the relationship between image restoration and network construction, independent of training data and using DL. The $med$ framework allows analysis of the deep prior by using four networks components (\textit{depth}, \textit{skip connections}, \textit{composition}, \textit{cascade}) whereas DIP \cite{Ulyanov2018CVPR} includes the investigation based on the two network components (\textit{depth} and \textit{skip connections}). The $med$ framework provides a more rigorous evaluation of the usefulness of skip connections compared to \cite{Ulyanov2018CVPR}. 
\item We also perform various image restoration tasks to show the quality of the image prior captured by the multi-level network architectures. We have achieved results comparable to the state-of-the-art methods for denoising, super-resolution, and inpainting with $x\%$ pixels drop despite experimenting with various high-capacity networks. We also observe a better flash no-flash based image construction when compared to \cite{Ulyanov2018CVPR}.
\end{itemize}

\section{Related work}
Image restoration aims to recover a good quality image from a corrupted observation. It is a useful preprocessing step for other problems, e.g., classification \cite{vidal2018ug}. Mao \textit{et al.} have shown image restoration using an $ed$ network with symmetric skip links between the layers of encoder and decoder \cite{mao2016image}. There are various proposals for the loss functions for the image restoration tasks, e.g., adversarial loss \cite{ledig2016photo}, perceptual loss  \cite{ledig2016photo}, or contextual loss \cite{mechrez2018contextual, mechrez2018learning}. In addition, Chang \textit{et al.} have proposed a single generic network for various image restoration tasks \cite{chang2017one}. However, the drawback to this line of work is that the restoration output could be biased toward a particular training dataset. 

Ulyanov \textit{et al.} showed that a randomly-initialized $ed$  network works as a hand-crafted prior for restoring images without training data \cite{Ulyanov2018CVPR}. Motivated by their approach, our learning-free framework only uses the handcrafted structure of the network for image restoration. However, unlike \cite{Ulyanov2018CVPR}, we explore how the network components directly influence various image restoration tasks. 

\section{Multi-level Encoder-Decoder Framework}\label{sec: networks}
In this section, we explain the multi-level encoder-decoder framework ($med$) and its major components. We shall also discuss an example construction of a multi-level encoder-decoder network and then provide a classification of the networks useful for our experiments. 

The $med$ is one of the general class of networks, where each network is a composition of encoder-decoder blocks as subnetworks. We address $med$ as a network $\mathcal{F}$ for devising a simpler explanation. The $med$ network $\mathcal{F}$ is a composition of two subnetworks, namely a \textit{generator} $G$ and an \textit{enhancer} $E$. The image restoration network $\mathcal{F}$ is defined in Eq.~\ref{eq: medf}.
\begin{dmath}\label{eq: medf}
\mathcal{F} = E \circ G
\end{dmath}
Here, the \textit{generator} and the \textit{enhancer} are either an encoder-decoder network ($ed$) or a composition of $ed$ networks. The encoders determine the abstract representation of the image features, which are used by the decoder for the reconstruction of the image. The composition of networks allows multiple sub-networks to learn image features from the down-sampled versions of the corrupted image. This would enforce the output of the \textit{generator} to be consistent across the multiple scales of the \textit{target} image\footnote{\textit{Target} image refers to the high-quality image whose corrupted observation $\hat{I}$ is given for restoration.} to improvise the quality of the image restoration. 

The multi-level encoder-decoder framework is motivated to model various network architectures by modifying the network components described in Subsection~\ref{ssec: networkComponents}. For example, let's suppose the \textit{generator} is a depth-$k$ $ed$ network. There are five network configurations obtained by modifying the skip connections, namely, \textit{Intra-skip}, \textit{Inter-skip encoder-encoder}, \textit{Inter-skip decoder-encoder}, \textit{No-skip}, and \textit{Full-skip} connections\footnote{In the supplementary material we have provided the details of different types of skip connections.}. There are two network configurations based on the cascading of the network input, \textit{i.e.}, network with \textit{cascade} or network without \textit{cascade}. There could be $(k-1)$ different \textit{generator}-\textit{enhancer} compositions for a depth-$k$ \textit{generator} network. We do not consider depth-$k$ \textit{enhancer} to reduce the model capacity. Finally, given a depth-$k$ $ed$ network as the \textit{generator}, the $med$ framework will allow $1 \times 5  \times 2 \times  (k-1)=10(k-1)$ different network structures. On the other hand, \cite{Ulyanov2018CVPR} will allow only two different network configurations (network with skip connections and without skip connections). Therefore, the generalization $med$ provides various networks to analyze the effects of network components on the quality of the image restoration. Technically, the $med$  is a general framework to explore the nature of the mapping between the network parameter space and the natural image space.

\subsection{Network Components}\label{ssec: networkComponents} 
We focus on the following components to show how the network structure affects the image restoration output. (a) \textit{skip} connections, (b) \textit{depth} of the network, (c) \textit{cascading} of the \textit{network input} into the intermediate layers, and (d) \textit{composition} of two $ed$ networks. We describe each of these components as follows.

\noindent \textbf{(a) Skip connections.} The skip link between the layers $L_i$ and $L_j$, where $i$ and $j$ are the indices of the network layers with $i<j$, is made by concatenating the output of the layer $L_{j-1}$ with the output of the layer $L_i$ and then feeding into the layer $L_j$. We have provided the detailed classification of the skip connections in supplementary material. In Fig.~\ref{fig: skipType}(a) and Fig.~\ref{fig: skipType}(b), we have pictorially shown useful skip link configurations for the paper.

\noindent \textbf{(b) Depth of the network.} It is measured by the number of layers present in the network. Higher depth networks capture finer feature details. However, a very high depth could negatively influence the performance (Fig.~\ref{fig: denoisingOurs}).  There are two ways to increase network depth. First, by introducing a new layer into the encoder-decoder ($ed$) network. Second, by performing a composition of the two $ed$ networks. 

\noindent \textbf{(c) Cascading of network input (\textit{cascade}).} It is a procedure to successively down-sample the network input and then feed it into the intermediate layers of the network. Formally, to provide the network input at the intermediate layer $L$, we resize the network input and then concatenate it with the layer $L-1$. Next, we feed the resulting tensor into the layer $L$. Cascading of network inputs was also utilized by Chen \textit{et al.} \cite{chen2017photographic}. We use it to provide the image features into the \textit{enhancer} network (Fig.~\ref{fig: skipType}(c)). 

\noindent \textbf{(d) Composition of $ed$ networks (\textit{composition}).} The composition of two encoder-decoder networks is achieved by feeding the output of the first $ed$ network into the second $ed$ network. The composition of two $ed$ networks increases the network depth and the number of skip connections. The main objective of performing the network composition is to learn image features from the downsampled versions of the corrupted image.

\begin{figure}[!htb]
\begin{center}
\begin{subfigure}[b]{0.20\textwidth}
\begin{center}
\resizebox{0.55\linewidth}{!}{%
\includegraphics{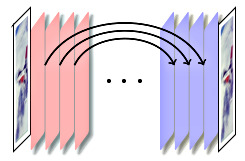}
} \end{center} \caption{\textit{Intra-skip.}}
\end{subfigure}
\begin{subfigure}[b]{0.25\textwidth}
\resizebox{0.7\linewidth}{!}{%
\includegraphics{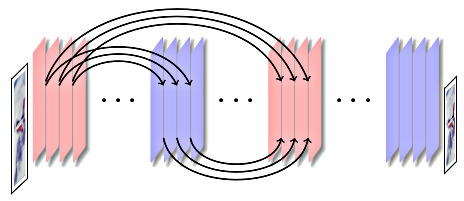}
} \caption{\textit{Full-skip.}}
\end{subfigure}
\begin{subfigure}[b]{0.35\textwidth}
\includegraphics[width=\textwidth]{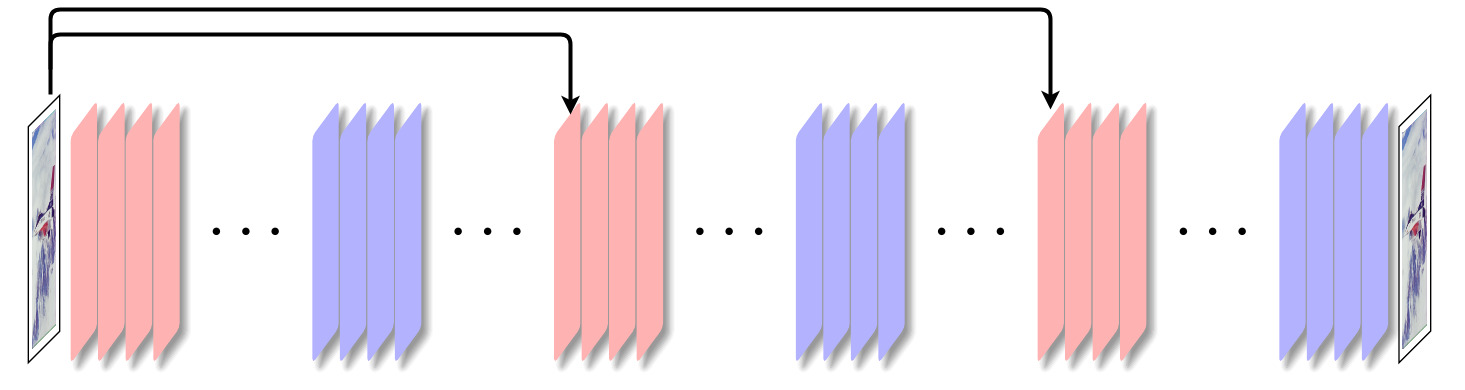} \caption{\textit{Cascading} of \textit{network input.}}
\end{subfigure}%
\end{center} \vspace*{-0.3cm} \caption{\textbf{Network components.} Layers of the encoder are in \textit{red} and layers of the decoder are in \textit{blue}. (a) \textit{Intra-skip}: the skip connections within EDS network. (b) \textit{Full-skip}: both the \textit{Intra-skip} connections and \textit{Inter-skip} connections are present. (c) Cascading of the network input.}\label{fig: skipType}
\end{figure}

\subsection{Multi-level Encoder-Decoder Network} 
\begin{figure*}[!htb]
\begin{center}
\resizebox{.75\linewidth}{!}{%
\includegraphics{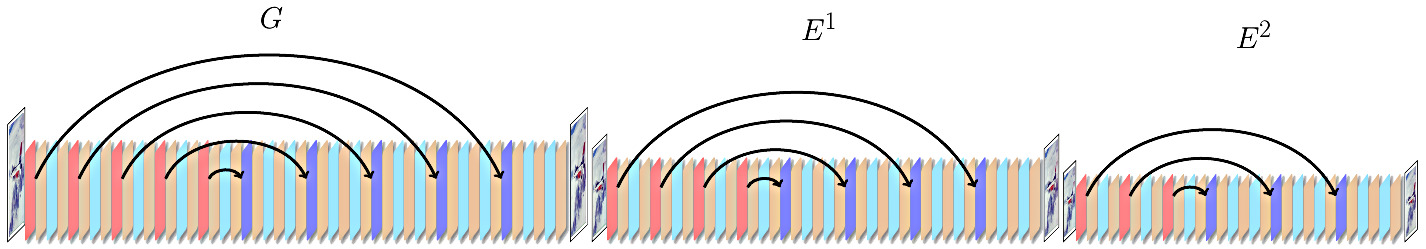}
}\hspace*{-0.5cm}
\caption{\textbf{Multi-level encoder-decoder network architecture.} An example construction of a three-level $med$ network. The \textit{generator} $G$ is an $ed$ network and \textit{enhancer} $E = E^1 \circ E^2$ is the composition of two $ed$ networks. There are skip connections within each $ed$ subnetwork. The layers are shown using colors as follows: \textbf{\texttt{{\color{cyan} Convolutional layer with stride =1}, {\color{red} Convolutional layer with stride=2}, {\color{brown} Batch Normalization}, and {\color{blue} Upsampling}}}. The subnetwork $G$ is a depth-5 $ed$ network, $E^1$ is a depth-4 $ed$ network, and $E^2$ is a depth-3 $ed$ network. }\label{fig: MEDS}
\end{center}
\end{figure*}

Here, we give an example construction of $med$ network $\mathcal{F}$. It is a three-level $ed$ network  where the \textit{generator} is the first $ed$ and the \textit{enhancer} is a composition of the other two $ed$ (Fig. \ref{fig: MEDS}). 
\begin{dmath}\label{eq: composition}
\mathcal{F} = E^2 \circ E^1 \circ G
\end{dmath}
In Eq.~\ref{eq: composition}, the subnetwork $G$ is the \textit{generator} and subnetwork $E^2 \circ E^1$ is the \textit{enhancer} $E$.

The networks $G$, $E^1$, and $E^2$ are defined as follows. $G:\mathbb{R}^{m \times n \times c} \rightarrow \mathbb{R}^{m \times n \times c}$, $E^1:\mathbb{R}^{\frac{m}{2} \times \frac{n}{2} \times c} \rightarrow \mathbb{R}^{\frac{m}{2} \times \frac{n}{2} \times c}$, and $E^2:\mathbb{R}^{\frac{m}{4} \times \frac{n}{4} \times c} \rightarrow \mathbb{R}^{\frac{m}{4} \times \frac{n}{4} \times c}$.

Here, $c$ is the number of channels (c is 3 for RGB images). The  \textit{generator} $G$ operates at $2\times$ the resolution of $E^1$ and $4\times$ the resolution of $E^2$. A resize operator $R$ is used to down-sample the  output of $G$ to feed into $E^1$ and down-sample the output of $E^1$ to feed into $E^2$. We have abstracted out $R$  in Eq.~\ref{eq: composition} for devising a simpler explanation. As described earlier, the \textit{enhancer} $E=E^2 \circ E^1$ is mainly used to improvise the output of the  \textit{generator} $G$ by making it consistent across different resolutions of the target images. 

\subsection{Network Classification}  
We have provided an example construction of a multi-level encoder-decoder network in Fig.~\ref{fig: MEDS}. Similarly, there are various other network architectures we can get by modifying the network components. We give a classification of $med$ networks useful for \emph{our methods} to analyze these network architectures. The $med$ network is classified based on skip links and cascading of the network input, as shown in Table~\ref{tab: medClass}. The network MED has no skip connections and MEDS has \textit{Intra-skip} connections (the character ``S'' in MEDS denotes the presence of skip connections). The network MEDSF has \textit{Full-skip} connections. Similarly, MEDC has cascading of network input without skip connections (the character ``C'' in MEDC denotes the cascading of network input). MEDSFC has cascading of network input with \textit{Full-skip} connections. We will use the networks given in Table~\ref{tab: medClass} for our experiments. For example, to see the effects of the decreasing skip links, one could perform image restoration with MEDSF, MEDS, and MED networks. 
\begin{table}[!htb] \setlength\extrarowheight{2pt}
\centering\resizebox{0.9\linewidth}{!}{%
\begin{tabular}{>{\centering\arraybackslash}p{50pt}  >{\centering\arraybackslash}p{40pt}  >{\centering\arraybackslash}p{40pt}  >{\centering\arraybackslash}p{50pt} }
&\textit{ No skip} & \textit{Intra-skip} & \textit{ Full-Skip} \\  \cline{2-4} 
\end{tabular}} \\\resizebox{0.9\linewidth}{!}{%
\begin{tabular}{>{\centering\arraybackslash}p{50pt} | >{\centering\arraybackslash}p{40pt} | >{\centering\arraybackslash}p{40pt} | >{\centering\arraybackslash}p{50pt} |}
\textit{Cascade} & MEDC & MEDSC & MEDSFC \\  \cline{2-4} 
\textit{No Cascade} & MED & MEDS & MEDSF \\  \cline{2-4}
\end{tabular}}\caption{\textbf{Classification of $med$ networks.} The classification is based on the following network components: skip-links and cascading of network input at intermediate layers. The graphical representations of the above network components are shown in Fig.~\ref{fig: skipType}.}\label{tab: medClass}
\end{table}

\section{Applications}\label{sec: applications}
In this section, we show the performance on the following image restoration tasks: super-resolution, denoising, inpainting, and flash no-flash. We provide the technical details of the experiments in the supplementary material.

The aim of image restoration is to reconstruct the image features given a corrupted image $\hat{I}$. The image $\hat{I}$ is computed by adding noise or blur or downsampling the target image $I$. Ulyanov \textit{et al.} formulated the image restoration problem to the setting of DL based learning-free framework  \cite{Ulyanov2018CVPR}. The image restoration framework is as follows.
\begin{dmath}\label{eq: imagePrior}
\begin{aligned}
\theta^* = & {\underset {\theta}{\operatorname {arg\,\text{min}}}} \; \mathcal{L}\big( \mathcal{F}_{\theta}(\hat{z}), \hat{I} \big); 
\end{aligned}
\end{dmath}
Here, $\mathcal{L}$ is the loss function and $\mathcal{F}$ is a network with parameters denoted by $\theta$ and the network input $z$ is prepared from the corrupted image $\hat{I}$. The loss function in the Eq.~\ref{eq: imagePrior} is a general definition. We now discuss how to perform various image restoration tasks. \\

\noindent \textbf{Denoising.}\label{sssec: denoising}
Denoising aims to reduce noise and recover the clean image where the learning process is assisted only by the corrupted image. Consider a noisy image $\hat{I}$. Let $d_1=\mathcal{D}(\frac{1}{2},\hat{I})$ and $d_2=\mathcal{D}(\frac{1}{4},\hat{I})$ be the down-sampled versions of the image $\hat{I}$. Our approaches are based on the following property of a natural image:  patch recurrence within and across multiple scales. Using this property, one could say that the down-sampled corrupted image contains some of the image features. To make the best use of the property above, our multi-scale loss $\mathcal{L}\big( \mathcal{F}_\theta(z), \hat{I} \big)$ (Eq.~\ref{eq: imagePrior}) for denoising is defined in Eq.~\ref{eq: denoisingLoss}. 
\begin{dmath}\label{eq: denoisingLoss}
\begin{split}
\theta^* = & \; {\underset {\theta}{\operatorname {arg\,\text{min}} }} \; \lambda_1 \|G_{\theta}(z) - \hat{I} \| \\
& + \lambda_2  \| E^1_{\theta}(z) - d_1 \| + \lambda_3  \| E^2_{\theta}(z) - d_2 \|
\end{split}
\end{dmath}
Here, $\mathcal{F} = E^2 \circ E^1 \circ G$ (Eq.~\ref{eq: composition}). Loss function in Eq.~\ref{eq: denoisingLoss} enforces the output of the \textit{generator} to be consistent across the multiple resolutions of the target image. Stated differently, the network performs image restoration at multiple resolutions. Intuitively, achieving restoration at multiple scales is more challenging than at a single scale. Therefore, we expect that solving a harder problem could help in learning a better image prior \cite{kim2018learning}.  The image prior is implicitly captured by the network which is required to restore the image features \cite{Ulyanov2018CVPR}.

\begin{figure}
\begin{subfigure}{0.75in}\captionsetup{justification=centering}\begin{center}
\includegraphics[width=\textwidth]{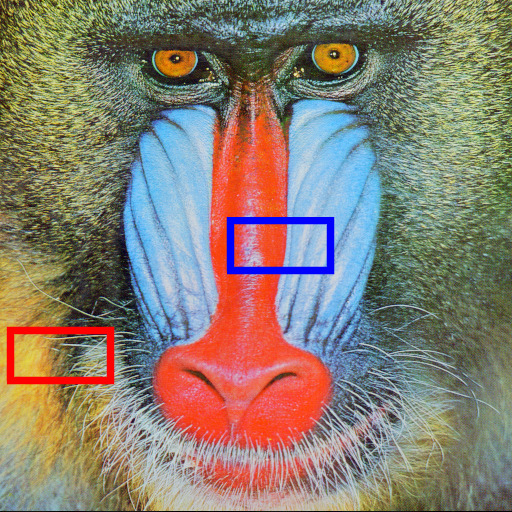} \\
\begin{subfigure}{0.37in}
\includegraphics[width=\textwidth, cfbox=blue 0.1pt 0.1pt]{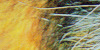}
\end{subfigure}%
\begin{subfigure}{0.37in}
\includegraphics[width=\textwidth, cfbox=red 0.1pt 0.1pt]{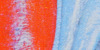}
\end{subfigure}
\end{center} \vspace*{-0.3cm}  \caption{Original, \\image} \end{subfigure}
\begin{subfigure}{0.75in}\captionsetup{justification=centering}\begin{center}
\includegraphics[width=\textwidth]{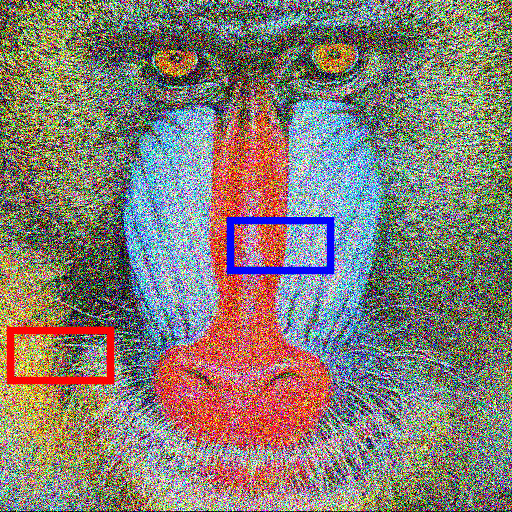} \\
\begin{subfigure}{0.37in}
\includegraphics[width=\textwidth, cfbox=blue 0.1pt 0.1pt]{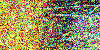}
\end{subfigure}%
\begin{subfigure}{0.37in}
\includegraphics[width=\textwidth, cfbox=red 0.1pt 0.1pt]{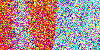}
\end{subfigure}
\end{center} \vspace*{-0.3cm}  \caption{Noisy, \\image} \end{subfigure}
\begin{subfigure}{0.75in}\captionsetup{justification=centering}\begin{center}
\includegraphics[width=\textwidth]{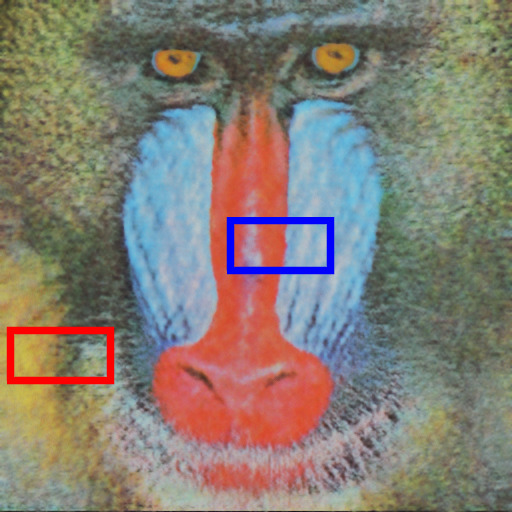} \\
\begin{subfigure}{0.37in}
\includegraphics[width=\textwidth, cfbox=blue 0.1pt 0.1pt]{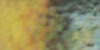}
\end{subfigure}%
\begin{subfigure}{0.37in}
\includegraphics[width=\textwidth, cfbox=red 0.1pt 0.1pt]{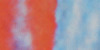}
\end{subfigure}
\end{center} \vspace*{-0.3cm}  \caption{DIP, \\(0.479, 18.65)} \end{subfigure}
\begin{subfigure}{0.75in}\captionsetup{justification=centering}\begin{center}
\includegraphics[width=\textwidth]{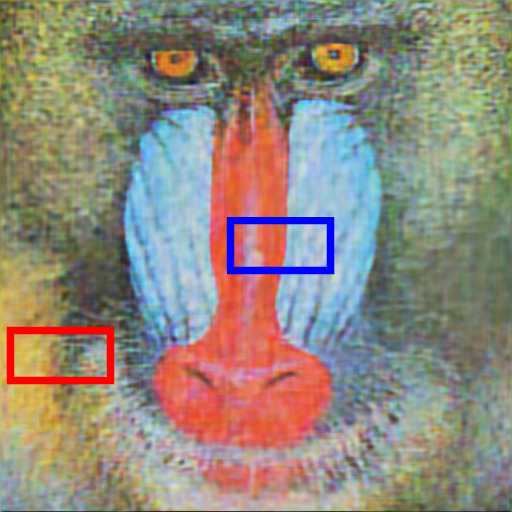} \\
\begin{subfigure}{0.37in}
\includegraphics[width=\textwidth, cfbox=blue 0.1pt 0.1pt]{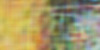}
\end{subfigure}%
\begin{subfigure}{0.37in}
\includegraphics[width=\textwidth, cfbox=red 0.1pt 0.1pt]{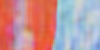}
\end{subfigure}
\end{center} \vspace*{-0.3cm}  \caption{Ours, \\(0.496, 18.39)} \end{subfigure}\vspace*{-0.1cm}
\caption{\textbf{Denoising}. A comparison between DIP \cite{Ulyanov2018CVPR} and our MEDSF for denoising with noise strength of $\sigma=100$ using the performance metric (SSIM, PSNR). }\label{fig: denoisingOurs}
\end{figure}

Denoising using our MEDSF is shown in Fig.~\ref{fig: denoisingOurs}. Our MEDSF achieves SSIM=0.72 whereas the baseline DIP \cite{Ulyanov2018CVPR} outputs a SSIM of 0.71 for a noise strength of $\sigma=100$. The PSNR values for our MEDSF is 20.95 and DIP outputs a PSNR of 21.36. In Fig.~\ref{fig: super}, we can observe that a higher PSNR value do not imply higher perceptual quality. 
We emphasize that the learning-free methods are sensitive to hyper-parameters\footnote{The learning-free methods are sensitive to hyper-parameters shown in Fig. 4 of the supplementary material and DIP  \cite{Ulyanov2018CVPR}.}. Therefore, the performance of DIP and our MEDSF could probably be further maximized by changing the hyper-parameters. 

\begin{figure}
\begin{center}
\resizebox{0.8\linewidth}{!}{%
\includegraphics[width=\textwidth]{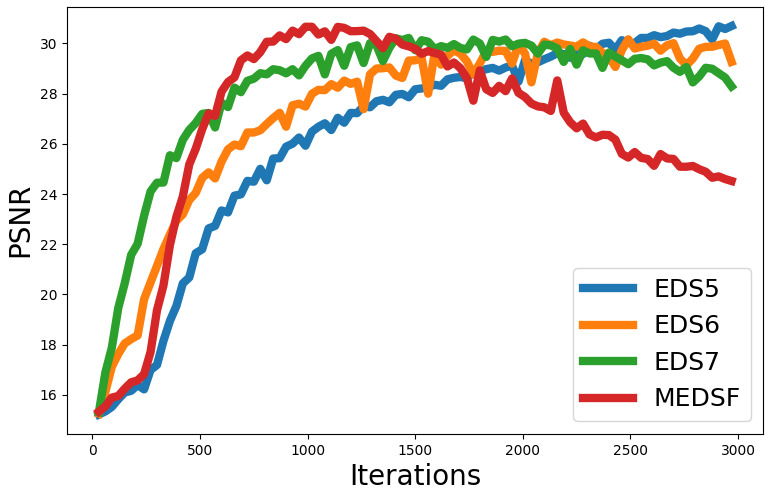}
}
\end{center}\vspace*{-0.5cm} \caption{\textbf{Network depth effects on denoising.} EDS5 is a depth-5 $ed$ network with skip connections (similarly for EDS6 and EDS7). The highest-depth network MEDSF converges faster.  EDS5 network (lower depth) achieves the highest PSNR value but converges the slowest. This shows that a higher model capacity does not necessarily lead to improved performance.}\label{fig: ed-depthEffects}
\end{figure}

\begin{figure*}[ht]\centering
\begin{subfigure}{0.20\textwidth}\captionsetup{justification=centering}\begin{center}
\includegraphics[width=\textwidth,cfbox=black 1pt 1pt]{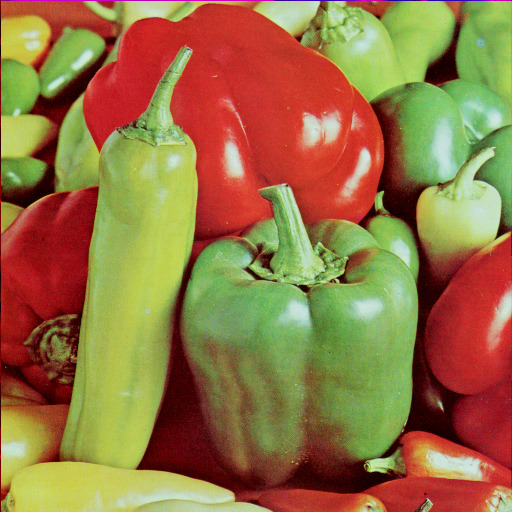}
\end{center} \vspace*{-0.3cm} \caption{High resolution image.} \end{subfigure} \hspace*{0.2cm}
\begin{subfigure}{0.20\textwidth}\captionsetup{justification=centering}\begin{center}
\includegraphics[width=\textwidth,cfbox=black 1pt 1pt]{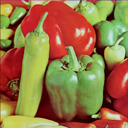}
\end{center} \vspace*{-0.3cm} \caption{Low resolution image.} \end{subfigure}  \hspace*{0.2cm}
\begin{subfigure}{0.20\textwidth}\captionsetup{justification=centering}\begin{center}
\includegraphics[width=\textwidth,cfbox=black 1pt 1pt]{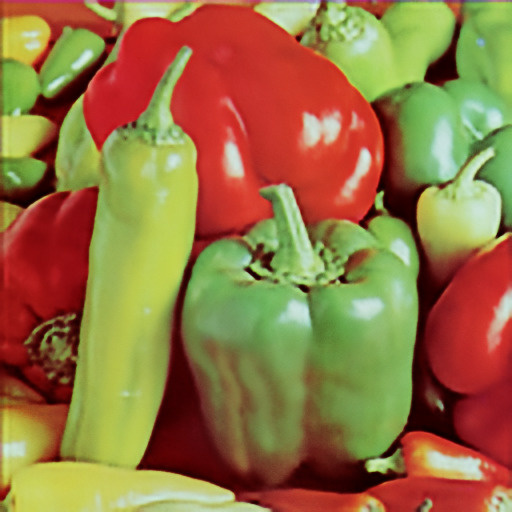} 
\end{center} \vspace*{-0.3cm} \caption{DIP \cite{Ulyanov2018CVPR}, (0.88, 28.2).} \end{subfigure}  \hspace*{0.2cm}
\begin{subfigure}{0.20\textwidth}\captionsetup{justification=centering}\begin{center}
\includegraphics[width=\textwidth,cfbox=black 1pt 1pt]{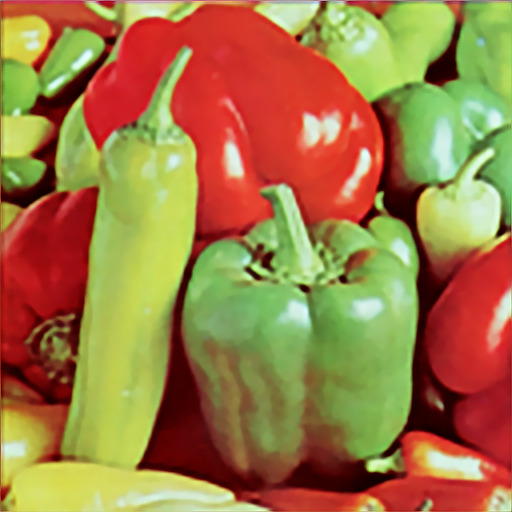}
\end{center} \vspace*{-0.3cm} \caption{MEDSF, (0.88, 25.43).} \end{subfigure} \vspace*{-0.2cm}
\caption{\textbf{$4\times$ Image super-resolution.} A qualitative comparision using performance metric (SSIM, PSNR). We can observe that a higher PSNR value does not imply a higher perceptual quality.}\label{fig: super}
\end{figure*}

In Fig.~\ref{fig: ed-depthEffects}, we can observe the effects of network depth on denoising. The network initially learns the global features from the corrupted image by minimizing the loss function defined in Eq.~\ref{eq: denoisingLoss}. Later, the network starts learning fine feature details which includes noise. Therefore, due to over learning, it produces noisy spots similar to the ones contained in the corrupted image. For example, MEDSF intermediate output at around 1000 iterations is the desired noise free image because it achieves the maximum PSNR. \\


\noindent \textbf{Super-resolution.} \label{sssec: super}
Given a \textit{low-resolution} (LR) image  $\hat{I}\in \mathbb{R}^{m\times n \times 3}$, and a scaling factor $t$, super-resolution aims to enhance the image quality and generate a \textit{high-resolution} (HR) image $I^H \in \mathbb{R}^{mt\times nt \times 3}$. We feed network input $z$ into $med$ network $\mathcal{F} = E^2 \circ E^1 \circ G$ and solve the following minimization problem given in Eq.~\ref{eq: superLoss}.
\begin{dmath}\label{eq: superLoss}
\begin{split}
\theta^* =  \; &{\underset {\theta}{\operatorname {arg\,\text{min}} }} \; \lambda_1 \| G_{\theta}(z) - u_0 \|  & \\
& + \lambda_2 \| E^1_{\theta}(z) - u_1 \| + \lambda_3 \| E^2_{\theta}(z) - \hat{I} \| &
\end{split} 
\end{dmath}
Here, $u_0 = \mathcal{U}(\hat{I},4)$ and $u_1=\mathcal{U}(\hat{I},2)$ are the up-sampled versions of the corrupted LR image $\hat{I}$. Eq.~\ref{eq: superLoss} determines the network parameter $\theta^*$ which minimizes the loss $\mathcal{L}\big( \mathcal{F}_\theta(z), \hat{I} \big)$.

Super-resolution achieved by Ulyanov \textit{et al.} in Deep Image Prior (DIP) is the state-of-the-art in DL-based learning-free methods to the best of our knowledge \cite{Ulyanov2018CVPR}. DIP does not use training samples to learn the image prior in contrast to the learning-based methods which benefit from the training data and \textit{adversarial} loss or \textit{perceptual} loss \cite{sajjadi2016enhancenet,ledig2016photo}. Thus, it lacks local level features in the output image. However, it is shown to output better images than various learning-free methods such as bicubic upsampling \cite{Ulyanov2018CVPR}. 

We achieved an average SSIM of 0.80, whereas DIP \cite{Ulyanov2018CVPR} achieved an average SSIM of 0.81 for $4\times$super-resolution. We obtained 24.48 as the average PSNR. Whereas DIP achieved an average PSNR of 25.14\footnote{RGB images in \href{https://github.com/jbhuang0604/SelfExSR/tree/master/data/Set14/image_SRF_4}{Set14} dataset had three channels and our $med$ network also outputs RGB images having three channels. However, super-resolution output of \href{https://dmitryulyanov.github.io/deep_image_prior}{DIP} \cite{Ulyanov2018CVPR} have images with four channels (including the \textit{alpha} channel). Therefore, to get a \emph{fair comparison}, we reproduced the DIP output before drawing the comparison.}. The perceptual quality of the generated images by the proposed approach is observed to be comparable to that of DIP (Fig.~\ref{fig: super}). \\

\noindent \textbf{Image inpainting.} \label{sssec: inpainting}
It involves computing missing pixel values in the corrupted image $\hat{I}$ using the corresponding binary mask $m\in \{ 0,1 \}^{k \times l}$. Inpainting has various applications such as removing undesirable objects and text in an image, restoring damaged paintings, and computing missing pixels lost during transmission. 

\begin{figure*}[!htb]
\begin{center}
\resizebox{0.8\linewidth}{!}{
\begin{subfigure}{0.22\textwidth}\captionsetup{justification=centering}\begin{center}
\includegraphics[width=\textwidth]{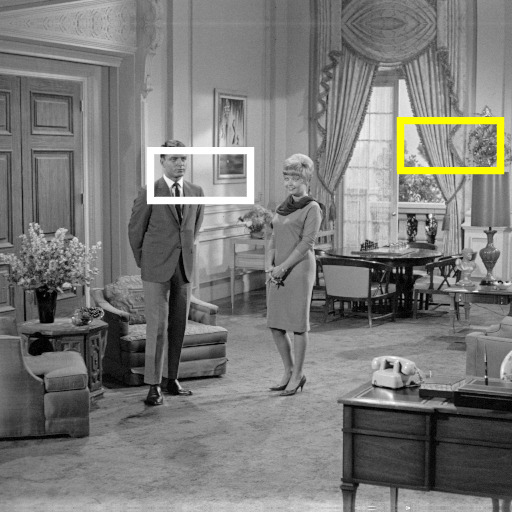} \\
\includegraphics[width=0.43\textwidth, cfbox=blue 0.1pt 0.1pt]{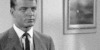}
\includegraphics[width=0.43\textwidth, cfbox=red 0.1pt 0.1pt]{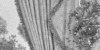}
\end{center} \vspace*{-0.3cm}  \caption{Original image} \end{subfigure}
\begin{subfigure}{0.22\textwidth}\captionsetup{justification=centering}\begin{center}
\includegraphics[width=\textwidth]{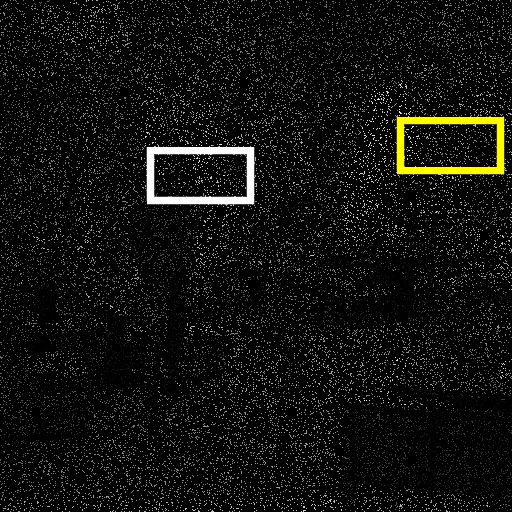} \\
\includegraphics[width=0.43\textwidth, cfbox=blue 0.1pt 0.1pt]{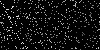}
\includegraphics[width=0.43\textwidth, cfbox=red 0.1pt 0.1pt]{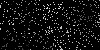}
\end{center} \vspace*{-0.3cm}  \caption{Corrupted image} \end{subfigure}
\begin{subfigure}{0.22\textwidth}\captionsetup{justification=centering}\begin{center}
\includegraphics[width=\textwidth]{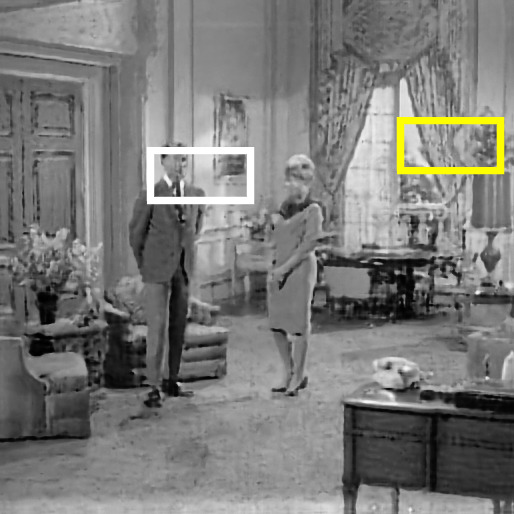} \\
\includegraphics[width=0.43\textwidth, cfbox=blue 0.1pt 0.1pt]{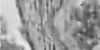}
\includegraphics[width=0.43\textwidth, cfbox=red 0.1pt 0.1pt]{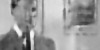}
\end{center} \vspace*{-0.3cm} \caption{DIP \cite{Ulyanov2018CVPR}, (0.85, 25.48)} \end{subfigure}
\begin{subfigure}{0.22\textwidth}\captionsetup{justification=centering}\begin{center}
\includegraphics[width=\textwidth]{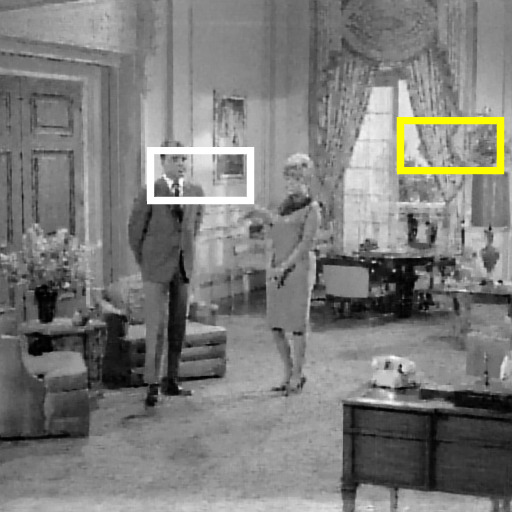} \\
\includegraphics[width=0.43\textwidth, cfbox=blue 0.1pt 0.1pt]{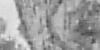}
\includegraphics[width=0.43\textwidth, cfbox=red 0.1pt 0.1pt]{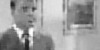}
\end{center} \vspace*{-0.3cm}  \caption{Ours, (0.85, 26.18)} \end{subfigure}
}\end{center} \vspace*{-0.5cm}
\caption{\textbf{Inpainting.} A comparision for restoration of $90\%$ missing pixels using performance metric (SSIM, PSNR).}\label{fig: restore2}
\end{figure*}

\begin{figure*}
\centering
\resizebox{0.8\linewidth}{!}{%
\begin{subfigure}{0.8in}\captionsetup{justification=centering}\begin{center}
\includegraphics[width=\textwidth]{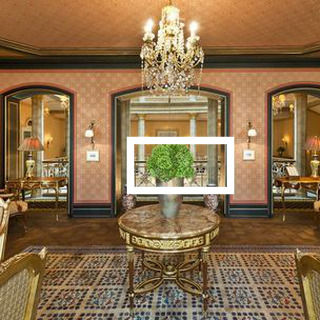} \\
\includegraphics[width=\textwidth, cfbox=white 0.1pt 0.1pt]{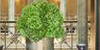}
\end{center} \vspace*{-0.3cm} \caption{Original\\image} \end{subfigure}
\begin{subfigure}{0.8in}\captionsetup{justification=centering}\begin{center}
\includegraphics[width=\textwidth]{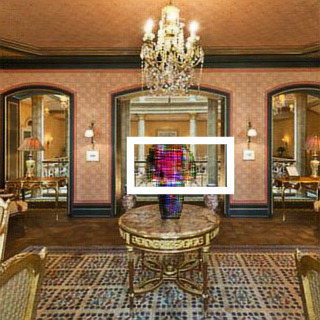} \\
\includegraphics[width=\textwidth, cfbox=white 0.1pt 0.1pt]{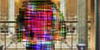}
\end{center} \vspace*{-0.3cm}  \caption{MEDSFC,\\24.87} \end{subfigure}
\begin{subfigure}{0.8in}\captionsetup{justification=centering}\begin{center}
\includegraphics[width=\textwidth]{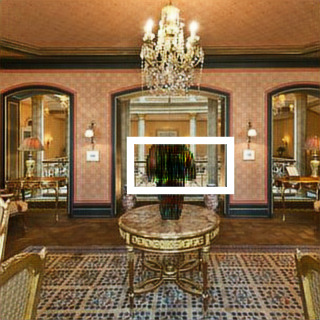} \\
\includegraphics[width=\textwidth, cfbox=white 0.1pt 0.1pt]{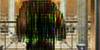}
\end{center} \vspace*{-0.3cm} \caption{MEDSF,\\23.20} \end{subfigure}
\begin{subfigure}{0.8in}\captionsetup{justification=centering}\begin{center}
\includegraphics[width=\textwidth]{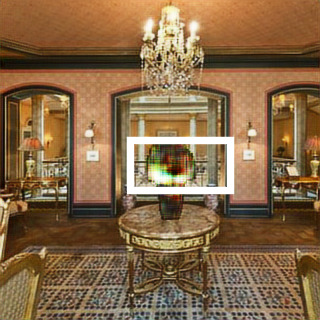} \\
\includegraphics[width=\textwidth, cfbox=white 0.1pt 0.1pt]{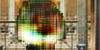}
\end{center} \vspace*{-0.3cm} \caption{MEDSC,\\24.28} \end{subfigure}
\begin{subfigure}{0.8in}\captionsetup{justification=centering}\begin{center}
\includegraphics[width=\textwidth]{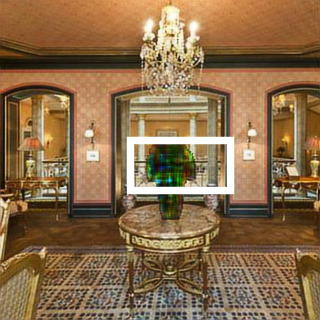} \\
\includegraphics[width=\textwidth, cfbox=white 0.1pt 0.1pt]{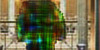}
\end{center} \vspace*{-0.3cm} \caption{MEDS,\\24.04} \end{subfigure}
\begin{subfigure}{0.8in}\captionsetup{justification=centering}\begin{center}
\includegraphics[width=\textwidth]{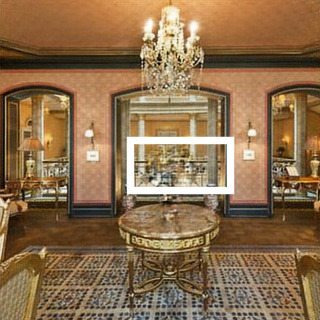} \\
\includegraphics[width=\textwidth, cfbox=white 0.1pt 0.1pt]{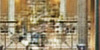}
\end{center} \vspace*{-0.3cm} \caption{MEDC,\\26.80} \end{subfigure}
\begin{subfigure}{0.8in}\captionsetup{justification=centering}\begin{center}
\includegraphics[width=\textwidth]{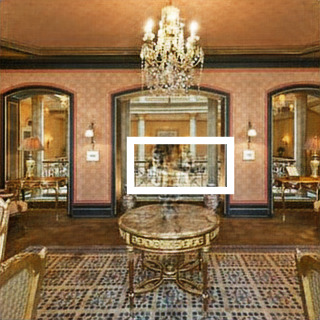} \\
\includegraphics[width=\textwidth, cfbox=white 0.1pt 0.1pt]{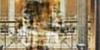}
\end{center} \vspace*{-0.3cm} \caption{MED,\\25.83} \end{subfigure}
} \vspace*{-0.2cm}
\caption{\textbf{Cascading of network input.} Effects of \textit{cascading} of the \textit{network input} in the intermediate layers of the network on removing an object from an image given in (a). The \textit{vase} present in (a) is removed using a white mask, and then inpainting is performed. Networks in (b) and (c) have the same set of skip links. Similarly, (d) and (e) have the same collection of skip links, and (f) and (g) do not have skip links. \textit{Cascading} of \textit{network input} is performed in (b), (d) and (f). (b) and (c) shows that cascading of network input into the intermediate layers of the network improves the performance. Similarly, we can observe that the cascading of network input performed better for other networks: (d) and (e), and (f) and (g). }\label{fig: objectRemove-cascade}
\end{figure*}

\begin{figure*}
\begin{center}
\resizebox{0.8\linewidth}{!}{%
\begin{subfigure}{0.8in}\captionsetup{justification=centering}\begin{center}
\includegraphics[width=\textwidth]{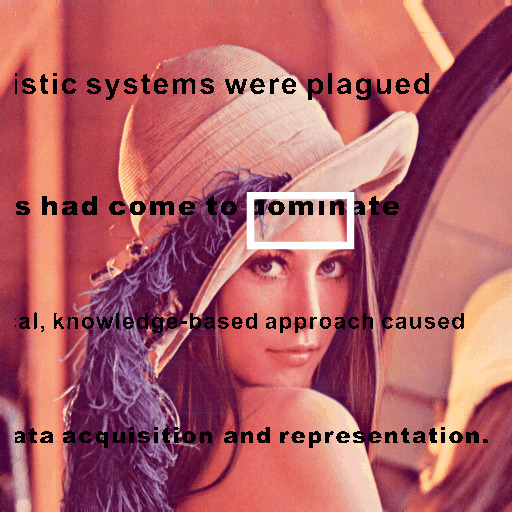} \\
\includegraphics[width=\textwidth, cfbox=white 0.1pt 0.1pt]{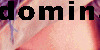}
\end{center} \vspace*{-0.3cm} \caption{Masked\\image} \end{subfigure}
\begin{subfigure}{0.85in}\captionsetup{justification=centering}\begin{center}
\includegraphics[width=\textwidth]{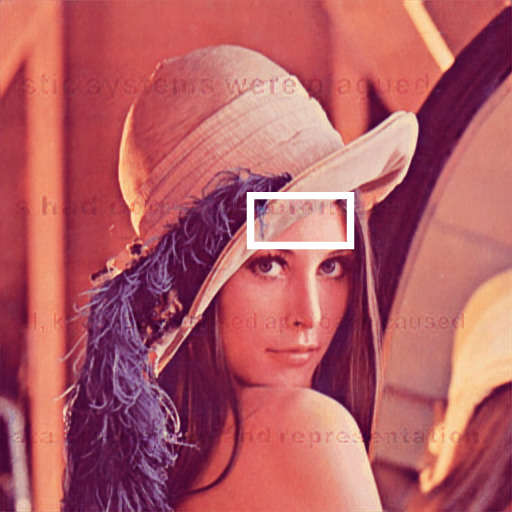} \\
\includegraphics[width=\textwidth, cfbox=white 0.1pt 0.1pt]{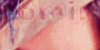}
\end{center} \vspace*{-0.3cm} \caption{MEDSF*,\\30.92} \end{subfigure}
\begin{subfigure}{0.85in}\captionsetup{justification=centering}\begin{center}
\includegraphics[width=\textwidth]{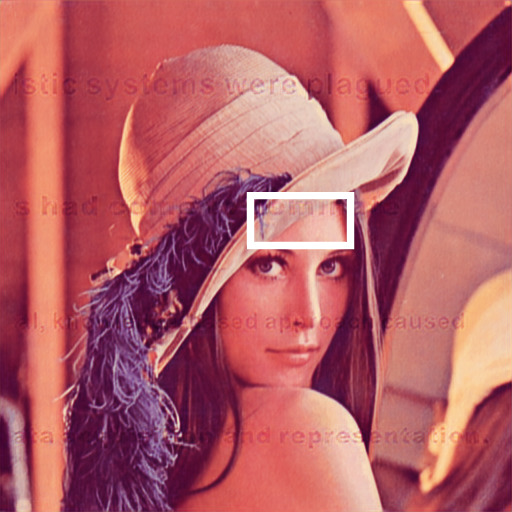} \\
\includegraphics[width=\textwidth, cfbox=white 0.1pt 0.1pt]{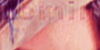}
\end{center} \vspace*{-0.3cm} \caption{MEDSF,\\30.40} \end{subfigure}
\begin{subfigure}{0.85in}\captionsetup{justification=centering}\begin{center}
\includegraphics[width=\textwidth]{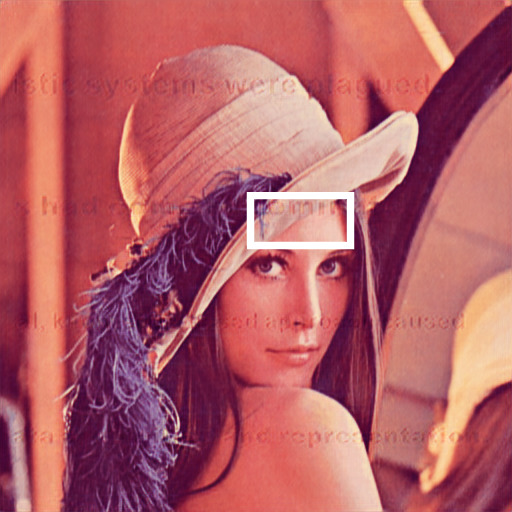} \\
\includegraphics[width=\textwidth, cfbox=white 0.1pt 0.1pt]{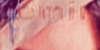}
\end{center} \vspace*{-0.3cm} \caption{MEDS*,\\31.05} \end{subfigure}
\begin{subfigure}{0.85in}\captionsetup{justification=centering}\begin{center}
\includegraphics[width=\textwidth]{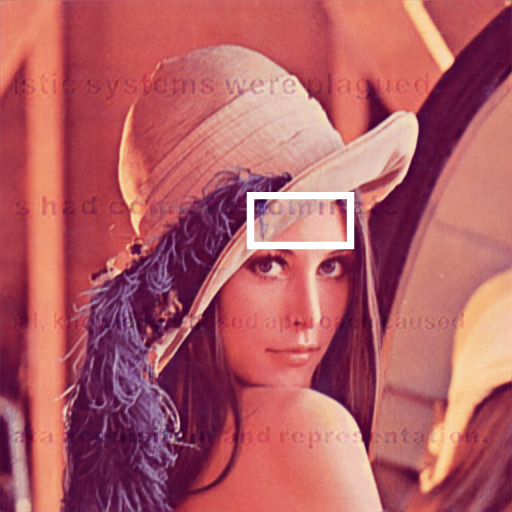} \\
\includegraphics[width=\textwidth, cfbox=white 0.1pt 0.1pt]{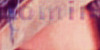}
\end{center} \vspace*{-0.3cm} \caption{MEDS,\\30.35} \end{subfigure}
\begin{subfigure}{0.85in}\captionsetup{justification=centering}\begin{center}
\includegraphics[width=\textwidth]{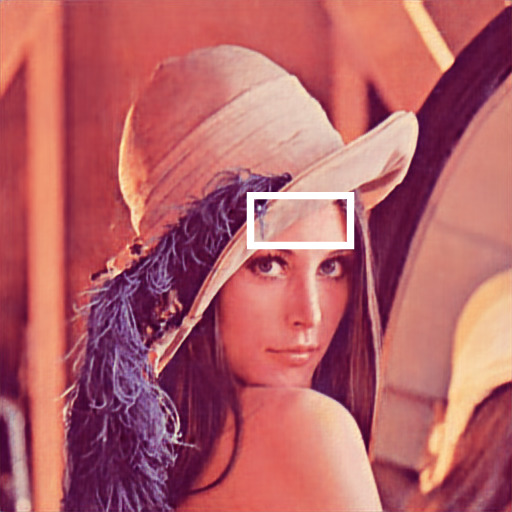} \\
\includegraphics[width=\textwidth, cfbox=white 0.1pt 0.1pt]{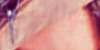}
\end{center} \vspace*{-0.3cm} \caption{MED*,\\30.18} \end{subfigure}
\begin{subfigure}{0.85in}\captionsetup{justification=centering}\begin{center}
\includegraphics[width=\textwidth]{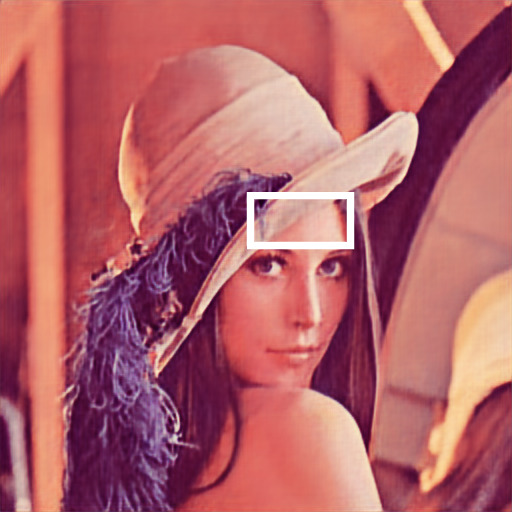} \\
\includegraphics[width=\textwidth, cfbox=white 0.1pt 0.1pt]{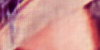}
\end{center} \vspace*{-0.3cm} \caption{MED,\\29.35} \end{subfigure}
} \end{center}\vspace*{-0.5cm}
\caption{\textbf{Composition of networks.} Effects of the composition of the $ed$ networks. MEDSF$^*$, MEDS$^*$, and MED$^*$ are two levels $ed$ networks. MED, MEDS, and  MEDSF are three level $ed$ networks (\textit{enhancer} is a composition of two $ed$ networks). (b) and (c) shows that the two-level full-skip network performed better than three levels of the full-skip network. Similarly, we can observe that the two-level $med$ network performed better for other networks: (d) and (e), and (f) and (g). }\label{fig: textRemove-composition}
\end{figure*}

Suppose $\mathcal{I}$ is the \textit{target} image and the corrupted image $\hat{I}$ is obtained using the mask $m$ as follows $\hat{I} = \mathcal{I} \odot m$, where $\odot$ is the Hadamard product. Let $d_1=\mathcal{D}(\frac{1}{2},\hat{I})$ and $d_2=\mathcal{D}(\frac{1}{4},\hat{I})$ be the down-sampled versions of the corrupted image $\hat{I}$, and $m_1=\mathcal{D}(\frac{1}{2},m)$ and $m_2=\mathcal{D}(\frac{1}{4},m)$ be the down-sampled versions of the mask $m$. We solve the following minimization problem given in Eq.~\ref{eq: inpaintingLoss}.
\begin{dmath}\label{eq: inpaintingLoss}
\begin{aligned}
\theta^* = \; &  {\underset {\theta}{\operatorname {arg\,\text{min}} }} \;  \lambda_1 \| (G_{\theta}(z) - \hat{I}) \odot m \| \\
& + \lambda_2 \| (E^1_{\theta}(z) - d_1) \odot m_2 \| + \lambda_3 \| (E^2_{\theta}(z) - d_2) \odot m_3 \|
\end{aligned}
\end{dmath}
We show the following three inpainting tasks. (1) \emph{restoring missing pixels} lost by masking the target image with a randomly generated binary mask (Fig.~\ref{fig: restore2}), (2) \emph{region-inpainting} which includes painting a large region (Fig.~\ref{fig: objectRemove-cascade} and Fig.~\ref{fig: skip1}), and (3) \emph{removing text} superimposed on an image (Fig.~\ref{fig: textRemove-composition}).

Inpainting requires understanding the global context and the local structure of the target image \cite{yang2017high}. We believe that region-inpainting is the most challenging task because the information from the nearby pixels might not always be sufficient to complete the scene. 

We obtained 24.62 as the average PSNR and 0.86 as the average SSIM for inpainting with $90\%$ missing pixels. Whereas DIP \cite{Ulyanov2018CVPR} achieved an average PSNR of 25.05 and an average SSIM of 0.86. 
The perceptual quality of the generated images by the proposed approach is observed to be comparable to that of the other methods  (Fig.~\ref{fig: restore2}). \\

\begin{figure}[ht]
\begin{center} \resizebox{\linewidth}{!}{%
\begin{subfigure}{0.8in}\captionsetup{justification=centering}\begin{center}
\includegraphics[width=\textwidth]{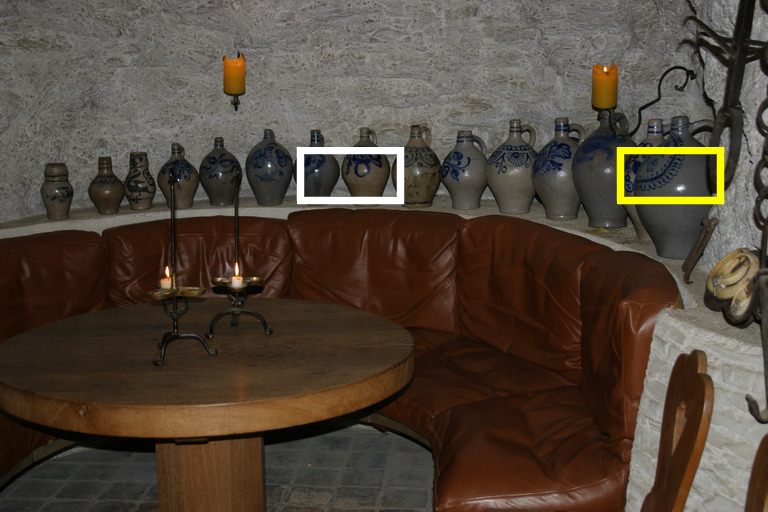} \\
\includegraphics[width=0.46\textwidth]{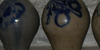}
\includegraphics[width=0.46\textwidth]{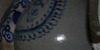}
\end{center} \vspace*{-0.3cm} \caption{{Flash image}} \end{subfigure}
\begin{subfigure}{0.8in}\captionsetup{justification=centering}\begin{center}
\includegraphics[width=\textwidth]{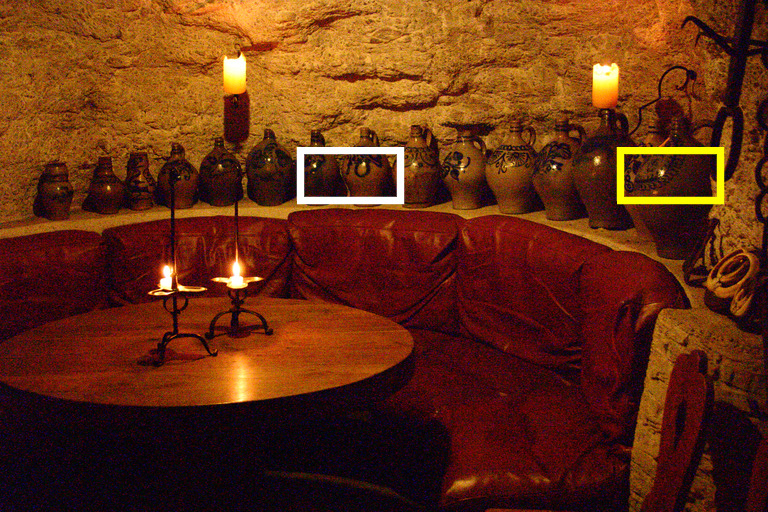} \\
\includegraphics[width=0.46\textwidth]{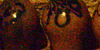}
\includegraphics[width=0.46\textwidth]{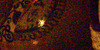}
\end{center} \vspace*{-0.3cm} \caption{{No-flash}} \end{subfigure}
\begin{subfigure}{0.8in}\captionsetup{justification=centering}\begin{center}
\includegraphics[width=\textwidth]{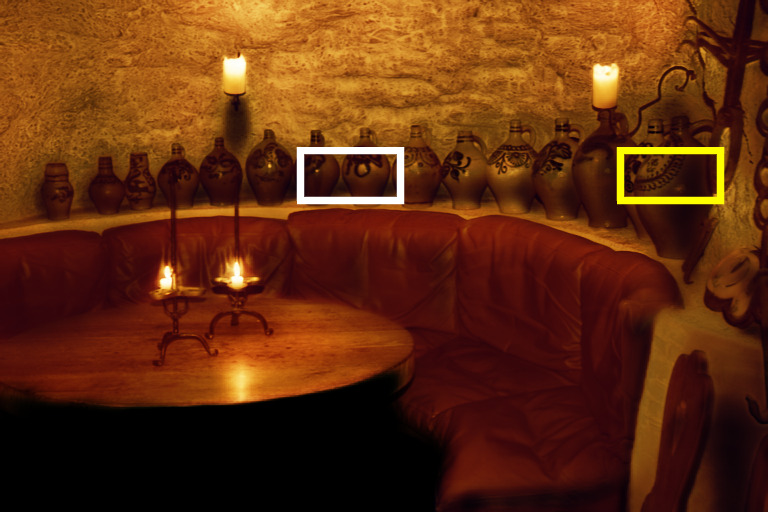} \\
\includegraphics[width=0.46\textwidth]{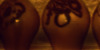}
\includegraphics[width=0.46\textwidth]{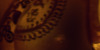}
\end{center}  \vspace*{-0.3cm} \caption{{DIP, 17.03}} \end{subfigure} 
\begin{subfigure}{0.8in}\captionsetup{justification=centering}\captionsetup{justification=centering} \begin{center}
\includegraphics[width=\textwidth]{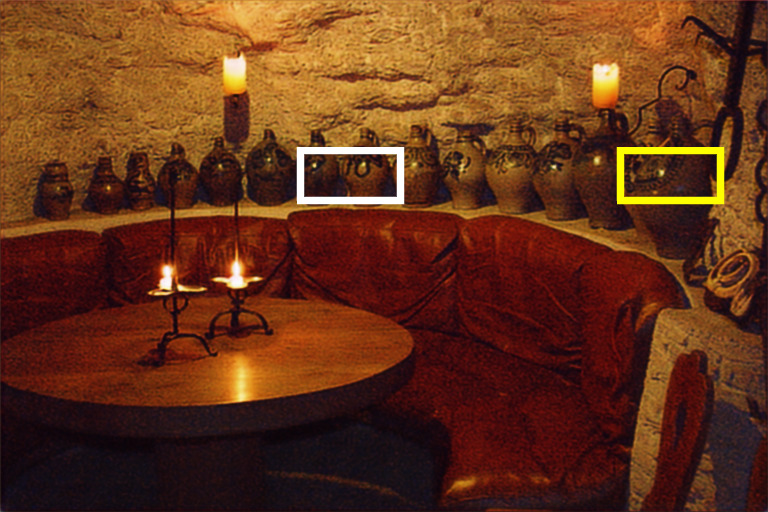} \\
\includegraphics[width=0.46\textwidth, cfbox=white 0.1pt 0.1pt]{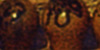}
\includegraphics[width=0.46\textwidth, cfbox=yellow 0.1pt 0.1pt]{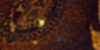}
\end{center} \vspace*{-0.3cm} \caption{{Ours, 18.54}} \end{subfigure} }
\end{center}\vspace*{-0.5cm}
\caption{\textbf{Flash-no flash reconstruction.} (a) Flash image. (b) No flash image. (c) DIP. (d) Ours MEDS.}\label{fig: flash}
\end{figure}%

\noindent \textbf{Flash No-flash.}\label{sssec: flash}
Given a pair of flash and no-flash images, the objective is to get a single high-quality image which incorporates details of the scene from the flash image and ambient illumination from the no-flash image \cite{petschnigg2004digital,eisemann2004flash}. The combined image helps to achieve denoising, white balancing, red-eye correction \cite{petschnigg2004digital}, foreground extraction \cite{sun2007flash}, and saliency detection \cite{he2014saliency}. 

Consider a pair $(I^F,I^{NF})$, where $I^F$ is a flash image and $I^{NF}$ is a no-flash image. The network input $z$ is prepared by concatenating $I^F$ and $I^{NF}$. Let $f_1=\mathcal{D}(\frac{1}{2},I^{NF})$ and $f_2=\mathcal{D}(\frac{1}{4},I^{NF})$ be the down-sampled versions of $I^{NF}$. We solve the optimization problem given in Eq.~\ref{eq: flash}.
\begin{dmath}\label{eq: flash}
\begin{aligned}
\theta^* = \; & {\underset {\theta}{\operatorname {arg\,\text{min}} }} \;  \lambda_1 \Big( \|G_{\theta}(z) - I^{NF} \| +  \| E^1_{\theta}(z) - f_1 \| \\ 
&  +  \| E^2_{\theta}(z) - f_2 \| \Big) + \lambda_2 \|G_{\theta}(z) - I^F  \|
\end{aligned}
\end{dmath}%
Here, $\lambda_1$ and $\lambda_2$ are the coefficients to control the image features from $I^{NF}$  and $I^F$. The flash no-flash output is shown in Fig.~\ref{fig: flash}. It is worth noting that our implementation of flash no-flash is more flexible in providing features from both flash and no-flash images using coefficients $\lambda_1$ and $\lambda_2$, unlike \cite{Ulyanov2018CVPR} (Fig. 12 of the supplementary material).

\section{Network Structures Effects on Restoration}
\label{sec: discussion}
\begin{figure}[!htb]
\centering \resizebox{\linewidth}{!}{%
\begin{subfigure}{0.20\textwidth}\captionsetup{justification=centering}\begin{center}
\includegraphics[width=\textwidth,cfbox=black 1pt 1pt]{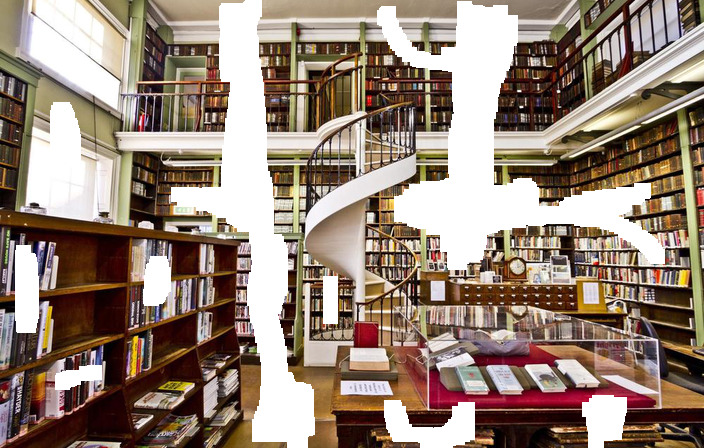}
\end{center} \vspace*{-0.3cm} \caption*{\large (a) Masked image.} \end{subfigure} \hspace*{0.05cm}
\begin{subfigure}{0.20\textwidth}\captionsetup{justification=centering}\begin{center}
\includegraphics[width=\textwidth,cfbox=black 1pt 1pt]{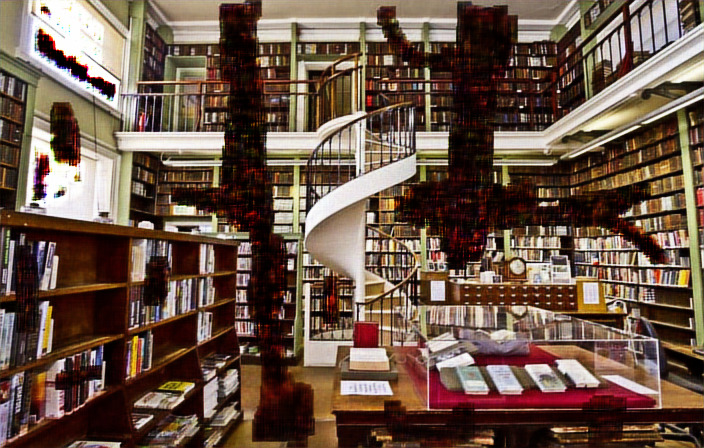}
\end{center} \vspace*{-0.3cm} \caption*{\large (b) MEDSF.} \end{subfigure} \hspace*{0.05cm}
\begin{subfigure}{0.20\textwidth}\captionsetup{justification=centering}\begin{center}
\includegraphics[width=\textwidth,cfbox=black 1pt 1pt]{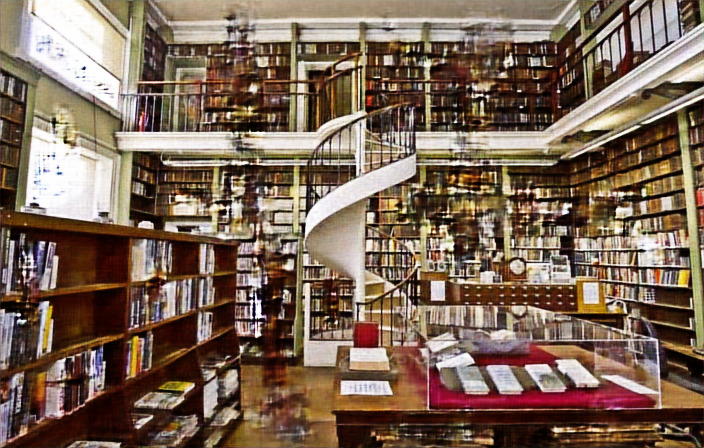}
\end{center} \vspace*{-0.3cm} \caption*{\large (c) MED.} \end{subfigure} \vspace*{-0.2cm}
}\caption{\textbf{Skip connections (I).} The network with skip links (MEDSF) does not perform well for \textit{region inpainting} compared to the network without skip connections.} \label{fig: skip1} 
\end{figure} %
\begin{figure}[!htb]
\resizebox{\linewidth}{!}{%
\begin{subfigure}{0.20\textwidth}\captionsetup{justification=centering}\begin{center}
\includegraphics[width=\textwidth,cfbox=black 1pt 1pt]{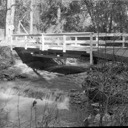}
\end{center} \vspace*{-0.3cm} \caption*{\large (a) LR image.} \end{subfigure} \hspace*{0.05cm}
\begin{subfigure}{0.20\textwidth}\captionsetup{justification=centering}\begin{center}
\includegraphics[width=\textwidth,cfbox=black 1pt 1pt]{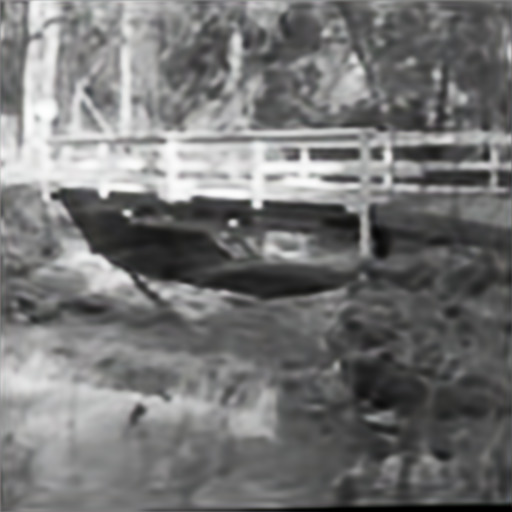}
\end{center} \vspace*{-0.3cm} \caption*{\large (b) MED, 0.63.} \end{subfigure} \hspace*{0.05cm}
\begin{subfigure}{0.20\textwidth}\captionsetup{justification=centering}\begin{center}
\includegraphics[width=\textwidth,cfbox=black 1pt 1pt]{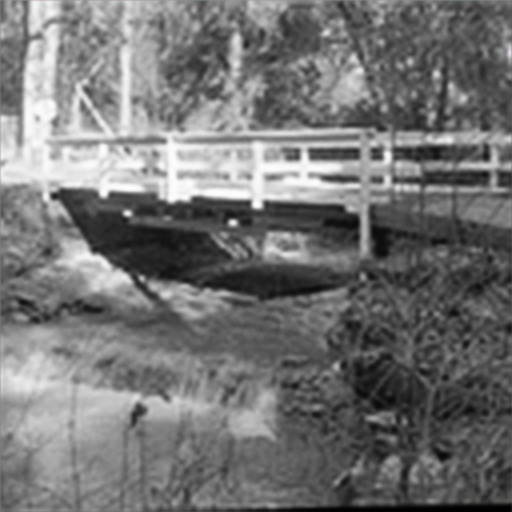}
\end{center} \vspace*{-0.3cm} \caption*{\large (c) MEDSF, 0.70.} \end{subfigure}} \vspace*{-0.2cm}
\caption{\textbf{Skip connections (II).} Skip links (MEDSF) improves 4$\times$ super-resolution as shown by SSIM.}\label{fig: skip2} \end{figure}
Here, we discuss the various aspects of the relation between the network construction and image restoration using the $med$ framework. Our choice of the multi-level architecture (a high capacity network) is motivated to illustrate the behavior of various network components (Sec.~\ref{sec: networks}). We emphasize that the image restoration quality from untrained networks is sensitive to hyper-parameters search \cite{Ulyanov2018CVPR}. We now discuss the results of the ablation studies that we have conducted. \\

\noindent \textbf{Effects of Skip links.} Skip connections have shown \textit{adverse} effects on inpainting, see Fig.~\ref{fig: objectRemove-cascade}, Fig.~\ref{fig: textRemove-composition}, and Fig.~\ref{fig: skip1} (the number of skip connections in the above figures decreases from left to right). Our interpretation of the adverse effects is as follows. The layers of \textit{encoder} have under-developed regions and their pixel values are close to that of the mask (either zero or one). The skip connections pass such intermediate representation to the decoder, which leads to reconstruction bias. Therefore, output images have pixel values that are close to the mask. \\

\noindent \textbf{Effects of Depth. } In Fig.~\ref{fig: ed-depthEffects}, we observe a higher the depth network converges faster because it has a large number of parameters. However, a lower depth network EDS5 could achieve better restoration than the higher depth network MEDSF. There could be two major factors for the above result. First, higher depth network suffer from the \textit{vanishing gradient} problem which negatively influences the performance \cite{mao2016image, srivastava2015training}. Second, the increase in the number skip connections due to higher depth, influence the performance positively \cite{mao2016image}. We believe that the decrease in the PSNR value indicates that the negative influence of the network depth could have more impact compared to the performance enhancement we get from skip connections.  \\

\noindent \textbf{Effects of Cascading of network input (\textit{cascade}).} The \textit{cascade} and the \textit{skip connection} looks similar because they both provide image features to the intermediate layers of the network. However, they provide a different type of image features. \textit{Cascade} provides image features from the corrupted image. Whereas, \textit{skip connections} pass the image features from the intermediate layers of the network. Object removal (inpainting) is better achieved using \textit{cascade} (Fig.~\ref{fig: objectRemove-cascade}). Whereas providing image features using skip connections have shown \textit{adverse} effects for inpainting (Fig.~\ref{fig: skip1}). This could be because of the image features captured at the intermediate layers of the network are less interpretable than the features present in the corrupted image. \vspace*{0.3cm} \\

\begin{table} \setlength\extrarowheight{2pt} \centering \resizebox{0.85\linewidth}{!}{%
\begin{tabular}{>{\centering\arraybackslash}p{80pt} >{\centering\arraybackslash}p{50pt} >{\centering\arraybackslash}p{50pt}}
& DIP \cite{Ulyanov2018CVPR} & $med$ (Ours) \\ 
\end{tabular}} 
\\ \resizebox{0.85\linewidth}{!}{%
\begin{tabular}{>{\centering\arraybackslash}p{80pt} | >{\centering\arraybackslash}p{50pt} | >{\centering\arraybackslash}p{50pt} |}  \cline{2-3}
\hspace*{1cm} Depth & \cmark & \cmark \\  \cline{2-3}
\hspace*{0.7cm} Skip-links & \cmark & \cmark \\  \cline{2-3}
Composition of $ed$ & \xmark & \cmark \\  \cline{2-3}
Cascading of input & \xmark & \cmark \\  \cline{2-3}
\end{tabular} } 
\caption{Network components to investigate the influence of the network structures for image restoration tasks.}\label{tab: allResults2}
\end{table}  \vspace*{-0.5cm}
\begin{table} \setlength\extrarowheight{2pt}\resizebox{0.85\linewidth}{!}{%
\begin{tabular}{>{\centering\arraybackslash}p{55pt} >{\centering\arraybackslash}p{70pt}  >{\centering\arraybackslash}p{70pt}}
& PSNR  & SSIM\\ 
\end{tabular} }\\\resizebox{0.85\linewidth}{!}{%
\begin{tabular}{>{\centering\arraybackslash}p{45pt} | >{\centering\arraybackslash}p{40pt} | >{\centering\arraybackslash}p{30pt} |  >{\centering\arraybackslash}p{40pt} | >{\centering\arraybackslash}p{30pt} | } \cline{2-5} 
& DIP \cite{Ulyanov2018CVPR}& Ours & DIP \cite{Ulyanov2018CVPR}& Ours \\  \cline{2-5} 
Denoising & 21.36 & 20.95 & 0.71 & 0.72 \\  \cline{2-5} 
Inpainting & 25.05 & 24.62 & 0.86 & 0.86 \\  \cline{2-5} 
SISR  & 25.14 & 24.48 & 0.81 & 0.80 \\  \cline{2-5} 
\end{tabular}} \caption{A quantitative comparison for denoising, inpainting, and single image super-resolution (SISR) using average PSNR and SSIM.  We provide the visual comparison of generated images in the supplementary material. The perceptual quality of the generated images is comparable to DIP \cite{Ulyanov2018CVPR} despite the $med$ network has a higher capacity to accommodate various network components (Table~\ref{tab: allResults2}).} \label{tab: allResults1}
\end{table}

\noindent \textbf{Effects of Composition of $ed$ networks.} The two-level $med$ network performed better than a three level $med$ network for text-removal from an image (Fig.~\ref{fig: textRemove-composition}). However, the performance difference is not very significant (less than one PSNR). A network composition increases the network depth and the number of skip connections. Therefore, a three-level $med$ could have more influence on restoration from skip connections compared to a two-level $med$ network. Similarly, a three-level $med$ could also increase the effects of \textit{vanishing gradients} due to the higher depth.  The composition of the networks shows the combined effects of increasing depth and skip connections.  

\section{Conclusion}
We have shown interesting aspects of the relationship between image restoration and network construction. Our methods are unsupervised and they only use the corrupted image for restoration instead of using any training data. Therefore, we believe that it does not produce a biased output unlike learning-based methods, e.g., model collapse \cite{salimans2016improved}. We feel it is a challenging experimental setup compared to supervised learning setup because the network is not learning image features by the pairs of low and high-quality images. The challenge is the limited contextual understanding due to the lack of feature learning from the training data. 

Our $med$ framework is a generalization of DIP \cite{Ulyanov2018CVPR}. This generalization is novel because it incorporates various network components and an \textit{enhancer} network. The $med$ framework is more expressive in terms of casting different network structures to perform the ablation studies for various aspects of the network (Table~\ref{tab: allResults2}). We also discuss image restoration task specific network structures that perform comparably to the state-of-the-art methods (Table~\ref{tab: allResults1}). 

The major components of the restoration framework are the network and the loss function (Eq.~\ref{eq: imagePrior}). We have shown analysis using various network structures and MSE loss\footnote{In Fig. 13 of the supplementary material, we show that MSE performed better than \textit{contextual loss} \cite{mechrez2018contextual}.}. The study of MSE loss is useful as it is used in other image restoration methods. For example, MSE with adversarial loss in \cite{ledig2017photo, shocher2018internal} and MSE with contextual loss in \cite{mechrez2018learning}. 

We observed that some network components do not enhance the restoration quality. For example, a network with skip links does not perform well for inpainting. Therefore, the experiments on a network with skip connections for inpainting will not be efficient. Wang \textit{et al.} have used skip connections for video inpainting \cite{wang2018video}. However, their approach is in the supervised learning setup, unlike our unsupervised setup. We believe that there are similarities in both setups. For example, if a network component is negatively influencing the image prior learning from the corrupted image (unsupervised setup), then it should also negatively influence the learning from the multiple images of training data (supervised setup). We propose as future work to study our restoration framework in the supervised learning setup. 

{\small
\bibliographystyle{ieee}
\bibliography{egbib}
}

\end{document}